\documentclass[12pt]{article}
\usepackage{mathrsfs,amssymb}
\usepackage{graphicx}
\usepackage{hyperref}

\begin{document}

\title{\bf Two-field K\"ahler moduli inflation on large volume moduli stabilization}

\author{{\sc Huan-Xiong Yang$^{\heartsuit,\spadesuit}$}\thanks{hyang@ustc.edu.cn},
{\sc Hong-Liang Ma$^{\heartsuit}$}\thanks{hlma@mail.ustc.edu.cn} \\
{\it $^{\heartsuit}$Interdisciplinary Center for Theoretical Study}, \\
{\it University of Science and Technology of China}, \\
{\it Hefei, 200026, P. R. China} \\
{\it $^{\spadesuit}$Kavli Institute for Theoretical Physics China}, \\
{\it CAS, Beijing 100190, China}
}

\date{\today}

\maketitle

\begin{abstract}
In this paper we present a two-field inflation model, which distinguishes itself with
a non-canonical kinetic lagrangian and comes from the large volume approach to
the moduli stabilization in flux compactification of type IIB superstring on a Calabi-Yau
orientifold of $h^{(1,2)} > h^{(1,1)}\geq 4$. The K\"ahler moduli are classified as volume
modulus, heavy moduli and two light moduli. The axion-dilaton, complex structure moduli
and all heavy K\"ahler moduli including the volume modulus are frozen by nonperturbatively
corrected flux superpotential and the $\alpha^\prime$-corrected K\"ahler potential in the
large volume limit. The minimum of the scalar potential at which the heavy moduli are
stabilized provides the dominant potential energy for the survived light K\"ahler moduli.
We consider a simplified case where the axionic components in the light K\"ahler moduli are
further stabilized at the potential minimum and only the geometrical components are taken
as the scalar fields to drive an assisted-like inflation. For a certain range of moduli
stabilization parameters and inflation initial conditions, we obtain a nearly flat power
spectrum of the curvature perturbation, with $n_s\approx 0.96$ at Hubble-exit, and an
inflationary energy scale of $3 \times 10^{14}$ GeV. In our model, significant correlation
exists between the curvature and isocurvature perturbations on super-Hubble scales so that
at the end of inflation a great deal of the curvature power spectrum originates from this
correlation.
\end{abstract}

\maketitle

\section{Introduction}\label{sec:intro}

The development of top-down cosmology has encountered a research hot on studying
the superstring landscape for realizing the cosmological inflation.
This involves the search of metastable de Sitter-like string vacua at which
the ten-dimensional spacetime behaves effectively as having only four dimensions,
with six spatial extra dimensions compactified on a Calabi-Yau threefold and almost
all of the compactification moduli including the Calabi-Yau volume stabilized.
The advent of KKLT mechanism\cite{kklt} for moduli stabilization has opened lots of
possibilities to engineer inflation in string theory, using either the light closed
string moduli\cite{quevedo04, quevedo061, quevedo062, bond} or open string
moduli\cite{kklmmt} related to the position of a mobile D-brane in wrapped geometry
as the relevant scalar fields. Different stringy realization of cosmological
inflation may not be mutually incompatible. They may arise in different regions of
the landscape, among which we anticipate that there is at least one possibility
that realizes the inflationary evolution of our Universe. The different inflation
scenarios may also turn into one another, increasing the overall probability of inflation
from a single mechanism\cite{bond}.

In a reasonable inflationary scenario, the inflaton field has to be ensured to roll its
potential very slowly so that there are enough efoldings for acceleration. It is necessary
that, during the inflation epoch, the potential energy curve of the involved scalars
must be sufficiently flat and their effective masses have to be much smaller than
the Hubble parameter, $m^2 \ll H^2$. Scalar fields such as the brane inter-distances
in a general D-brane inflation model are conformally coupled to gravity, which enables
them to acquire squared masses $m^2\approx \frac{1}{12}R\sim H^2$ ($\eta$-problem).
To overcome this $\eta$-problem and have a slow-roll inflation in such a scenario,
severe and sometimes unnatural fine-tunings are indispensable\cite{kklmmt}. Alternatively,
no severe $\eta$-problem exists in the closed string moduli inflation
scenario\cite{quevedo04, quevedo061, quevedo062, bond} where the inflation is thought
of to be driven by some light closed string moduli associated with 4-cycle volumes of
the compact Calabi-Yau threefolds, particularly in the large volume limit\cite{quevedo051,
quevedo052}.

The first proposed closed string moduli inflation scenarios in Type IIB superstring theory
are the racetrack inflation model\cite{quevedo04} and the better racetrack inflation
model\cite{quevedo061}, where the heavy moduli are stabilized at a supersymmetric anti-de
Sitter F-term potential minimum and the potential is uplifted to have a de Sitter vacuum
for inflation in an uncontrollable manner unless some additional ingredients are added
to break  the supersymmetry spontaneously\cite{achucarro, quevedo071, wen, misra}.
In the subsequent K\"ahler inflation model\cite{quevedo062} and Roulette inflation
model\cite{bond}, the scalar potential for the moduli fields is calculated in the large
volume limit, where the leading $\alpha^\prime$-correction to the K\"ahler potential has
been taken into account so that the heavy moduli are stabilized at a supersymmetric broken
anti-de Sitter F-term potential minimum. Because this anti-de Sitter minimum is apart from
supersymmetry, it can be uplifted to a de Sitter vacuum by taking into account some D-term
corrections to the scalar potential\cite{choi, dealwis}. Moreover, after heavy moduli are
stabilized in the large volume limit, the scalar potential for the survived light K\"ahler
moduli are exponentially flat, making such approaches very promising for engineering
assisted-like inflation\cite{lms}.

In string theory cosmology scenario, the early evolution of our universe is essentially
drived by multiple scalar fields. When superstring is compactified on a Calabi-Yau
threefold to yield a four-dimensional effective ${\mathcal N}=1$ supergravity theory,
lots of moduli fields emerge. It has been conjectured, however, that increasing the number
of K\"ahler moduli fields does probably make the inflation be easier to
achieve\cite{quevedo061}. Moreover, the presence of multiple scalar fields during inflation
can lead to quite different inflationary dynamics such as the detectable non-Gaussianity
in primordial density perturbations on super-Hubble scales and residual isocurvature
fluctuations after inflation\cite{wands} that might appear unnatural in a single field model.

Our aim in the present paper is to build a two-field inflation model from the large volume
limit of type IIB superstring with 3-form fluxes compactified on a Calabi-Yau orientifold
of $h^{(1,2)} >h^{(1,1)}\geq 4$ and discuss its implications on COBE spectrum of the linear
curvature perturbation. For that we first investigate the possibility of realizing better
racetrack inflation in the large volume approach. We divide the K\"ahler moduli into volume
modulus, $h^{(1,1)}-3$ heavy moduli and two light moduli. The scalar potential is dominated
by its F-term contribution, and is corrected by a volume-modulus dependent uplifting.
The axion-dilaton, complex structure moduli and all heavy K\"ahler moduli including
the volume modulus are stabilized, in the large volume limit, to a de Sitter-like minimum
of the potential, whereas the two light K\"ahler moduli remain dynamical during the moduli
stabilization process. These two light K\"ahler moduli are supposed to be the scalar fields
driving the inflationary expansion of the 4-dimensional universe. The de Sitter minimum of
the dominant potential does not only stabilize the heavy moduli by giving them masses, it
governs also the evolution of two survived light moduli. The exponential dependence of the
nonperturbatively corrected superpotential upon the K\"ahler moduli guarantees the flatness
of the scalar potential for two light K\"ahler moduli. If the axionic components of the
light moduli are dynamical, this is a large volume version of the better racetrack
model\cite{quevedo061} in which four scalar fields are involved. In the present paper we
choose to stabilize the axionic components at the potential minimum, a two-field
assisted-like model\cite{lms} results in instead. In some sense, this is an extension of
the single field inflationary model\cite{quevedo062} to two-field case, to which we
calculate the linear perturbations in a strict multi-field approach. By assigning some
suitable model dependent parameters, we obtain the COBE normalization favored power
spectrum of the linear curvature perturbation, and in particular, $n_s\approx 0.96$ at the
horizon exit.  Dependent upon the trajectory in phase space and thus the initial conditions
we choose, the isocurvature perturbation in this model could decay very slowly for most of
efolds on superhorizon scales. We have also observed that the power spectrum of curvature
perturbation on the super-Hubble scales has some remarkable departure from that of
adiabatic perturbation at the end of inflation, implying that the curvature fluctuation
after inflation does inevitably include significant contributions from isocurvature
perturbation.

The paper is organized as follows. In the second section, the theoretical setup is
introduced. We review briefly the large volume approach to the moduli stabilization
developed by Quevedo and his collaborators in \cite{quevedo051, quevedo052}
and establish our own two-field model. The third section is devoted to the investigation
of possible inflationary dynamics in our model. We first present the equations of motion
for the background fields and the linear perturbations, and then calculate the power
spectra of these perturbations and their correlation by numerical integration. The obtained
curvature power spectrum is nearly flat during most inflation and is characterized by
a  spectral index $n_s\approx 0.96$ at the claimed Hubble crossing. The final section
is our conclusions.

\section{Inflationary landscape in large volume approach}\label{sec:setup}

Inflation in string theory is always associated with the resolvability of the moduli
stabilization because the potential used to stabilize the heavy moduli might yield
unacceptably large masses for the inflation drivers\cite{kklmmt}. In type IIB superstring
compactified on a Calabi-Yau orientifold, the closed string moduli consist of the complex
structure moduli, dilaton field and the K\"ahler moduli. All complex structure moduli
including the dilaton field could belong to the heavy moduli and be stabilized at a minimum
of the effective potential if we introduce an imaginary self-dual 3-form flux into the
orientifold construction\cite{sethi, giddings}. The stabilization of the K\"ahler moduli
requires, according to KKLT mechanism\cite{kklt}, introducing the nonperturbative
contributions of the Euclid D3-branes or some wrapped D7-branes to the superpotential,
\begin{equation}\label{eq: 1}
W = \int G_3 \wedge \Omega + \sum_{i=1}^{h^{(1,1)}}A_i e^{-a_i T_i}
\end{equation}
where $T_i = \tau_i + i \theta_i$ with $\tau_i$ the $i$-th 4-cycle volume and $\theta_i$
the corresponding axions. The coefficients $A_i$ represent threshold corrections and are
independent of the K\"ahler moduli.

The K\"ahler potentials arising from type IIB superstring are of no-scale or approximately
no-scale\cite{louis},
\begin{equation}\label{eq: 2}
K = K_{cs} -2 \ln\left( {\mathcal V}
%\alpha (\tau_1^{3/2} - \sum_{i=2}^n \lambda_i \tau_i^{3/2})
+ \frac{\xi}{2}\right)
\end{equation}
satisfying
\begin{equation}\label{eq: 3}
K^{i\bar{j}}\partial_iK \partial_{\bar{j}}K -3 = \frac{3\xi}{4{\mathcal V}-\xi} \approx 0
\end{equation}
in the large volume limit ${\mathcal V}\rightarrow \infty$, where $\xi = - \frac{\zeta(3)
\chi(M)}{2(2\pi)^3 g_s^{3/2}}$, $\zeta(3)\approx 1.2$ and $\chi(M)=2(h^{(1,1)}-h^{(1,2)})$.
In (\ref{eq: 2}) the $\alpha^\prime$-corrections have been included. Combining with the
superpotential given in Eq.(\ref{eq: 1}),  we see that the K\"ahler moduli do only appear
exponentially in the F-term scalar potential\footnote{Throughout we take the units $M_{p}
= 1/\sqrt{8\pi G}=1$.},
\begin{equation}\label{eq: 4}
\begin{array}{lll}
V_F & = & e^K (K^{i\bar{j}}D_i K D_{\bar{j}}K  - 3 |W|^2) \\
& = & e^K K^{i \bar{j}}\left[a_i A_i a_j \bar{A}_j e^{-a_iT_i -a_j \bar{T}_j}
-(K_i W a_j \bar{A}_j e^{-a_j \bar{T}_j} + c.c.)\right]
\end{array}
\end{equation}
where $D_i K = \partial_i K W + \partial_i W$. The uplifting correction to the scalar
potential, which has several possible sources and is necessary to fine-tune the
cosmological constant, depends on moduli only through the overall volume,
\begin{equation}\label{eq: 5}
V_{uplift} \sim \frac{1}{{\mathcal V}^{\varrho}}
\end{equation}
where $\frac{4}{3}\leq \varrho \leq 3$. For concreteness, we assume $\varrho=2$ in the
present paper. It is remarkable that, in the presence of several K\"ahler moduli, the
variation of scalar potential along each $T_i$ direction is in general uncorrelated with
its magnitude. Besides, the potential is exponentially flat along some $T_i$ directions
as long as these K\"ahler moduli are sufficient large. Therefore, the large K\"ahler moduli
have much chance to become the scalar fields to drive the early inflationary evolution of
our Universe.

The Calabi-Yau orientifolds of Swiss-cheese type have been shown to be very useful in
string inflation engineering, whose volume can be formulated into a simplified form
as follows\cite{quevedo062, bond},
\begin{equation}\label{eq: 6}
{\mathcal V} = \frac{\alpha_0}{2\sqrt 2}\left[(T_1 + \bar{T}_1)^{3/2}
- \sum_{i=2}^n \lambda_i (T_i + \bar{T}_i)^{3/2}\right]
\end{equation}
Here the volume of the $i$-th 4-cycle is denoted by $\tau_i=\Re{T}_i$, among which $\tau_1$
controls the overall volume and $\tau_2, \cdots, \tau_n$ [$n=h^{(1,1)}$] are blow-ups whose
only non-vanishing triple intersections are with themselves. We stabilize the dilaton and
complex structure moduli with fluxes, following the plausible KKLT procedure\cite{kklt}.
After that, the superpotential (\ref{eq: 1}) is reduced to be only K\"ahler moduli
dependent, $W = W_0 + \sum_{i=1}^n A_i e^{-a_i T_i}$, where $a_i = 2\pi/{N}$ and for
simplicity $W_0$ is assumed to be real.

We establish our formalism in the large volume limit, where ${\mathcal V}\rightarrow
\infty$, $\tau_1 \approx {\mathcal V}^{2/3}\rightarrow \infty$, $a_i\tau_i \sim
\ln{\mathcal V}$ for $i=2, ~3, \cdots, ~n$ but $A_i\gg A_j$ for $i=2, ~3, \cdots,
~k$ and $j=k+1, ~k+2, \cdots, ~n$. Because $A_i$ in superpotential (\ref{eq: 1}) are
directly proportional to the squared masses of the corresponding K\"ahler moduli, $T_i$
are referred to be the heavy moduli when $i=2, ~3, \cdots, ~k$ and light moduli when
$i=k+1, ~k+2, \cdots, ~n$, respectively. In this limit, $e^K \approx 1/{\mathcal V}^2$,
the superpotential is approximately independent of $T_1$ which we will call the volume
modulus,
\begin{equation}\label{eq: 7}
W \approx W_0 + \sum_{i=2}^n A_i e^{-a_i T_i}
\end{equation}
Up to the magnitude of order $1/{\mathcal V}^3$, we can approximately write the scalar
potential as $V\approx V_{dom} + V_{corr}$, where the dominant part $V_{dom}$ consists
of the contributions of the volume modulus and the heavy moduli, and of the necessary
uplifting potential,
\begin{equation}\label{eq: 8}
\begin{array}{lll}
V_{dom} & = & \sum\limits_{i=2}^k \frac{8(a_i A_i)^2 \sqrt{\tau_i}}{3{\mathcal V}\lambda_i
\alpha_0} e^{-2a_i\tau_i}
+ \sum\limits_{i=2}^k \frac{4a_iA_iW_0\tau_i}{{\mathcal V}^2} e^{-a_i\tau_i}
\cos(a_i\theta_i) \\
&  & ~~ + \frac{3 \xi W_0^2}{4 {\mathcal V}^3}
+ \frac{\gamma W_0^2}{{\mathcal V}^2}
\end{array}
\end{equation}
The contributions of the light K\"ahler moduli to the scalar potential do only appear in
correction term $V_{corr}$,
\begin{equation}\label{eq: 9}
V_{corr} = \sum\limits_{i=k+1}^n \frac{8(a_i A_i)^2 \sqrt{\tau_i}}{3{\mathcal V}
\lambda_i \alpha_0} e^{-2a_i\tau_i}
+ \sum\limits_{i=k+1}^n \frac{4a_iA_iW_0\tau_i}{{\mathcal V}^2} e^{-a_i\tau_i}
\cos(a_i\theta_i)
\end{equation}
In Eq.(\ref{eq: 8}), we have parameterized the uplifting potential as $V_{uplift}
= \frac{\gamma W_0^2}{{\mathcal V}^2}$, with $\gamma$ a positive parameter.
There are terms not included in Eqs.(\ref{eq: 8}) and (\ref{eq: 9}), which are subleading.

In the large volume approach\cite{quevedo051, quevedo052}, the moduli stabilization
problem of the K\"ahler moduli is solved in the following two-step procedure\cite{holman}.
First, we stabilize the axions $\theta_i$ ($i = 2, 3, \cdots, k$) to the potential
minimum by setting $a_i\theta_i=\pi$. Second, we can find an approximate minimum of
the potential by letting $V_{dom}$ be flat along the directions of ${\mathcal V}$
and the heavy moduli $\tau_i$ ($i= 2, 3, \cdots, k$) in moduli space. The minimum
of the scalar potential turns out approximately to be,
\begin{equation}\label{eq: 10}
V_{min} \approx -\frac{3W_0^2}{2{\mathcal V}_f^3}\left[\sum\limits_{i=2}^k
\frac{\lambda_i \alpha_0}{a_i^{3/2}} (\ln {\mathcal V}_f -c_i)^{3/2} -\frac{\xi}{2}\right]
+ \frac{\gamma W_0^2}{{\mathcal V}_f^2}
\end{equation}
where ${\mathcal V}_f$ that stands for the stabilized Calabi-Yau volume is the solution
of the algebraic equation
\begin{equation}\label{eq: 11}
{\mathcal V} \approx - \frac{9\xi}{8\gamma} + \frac{9\alpha_0}{8\gamma}\sum\limits_{i=2}^k
\frac{\lambda_i \sqrt{(\ln {\mathcal V} -c_i)}}{a_i^{3/2}}[2\ln{\mathcal V} -(2c_i + 1)]
\end{equation}
with
\[
c_i = \ln\left(\frac{3\alpha_0 \lambda_i W_0}{4a_iA_i} \right)
\]
The heavy moduli will be stabilized at $a_i\tau_{i,min}\approx  \ln{\mathcal V}-c_i$.
Phenomenologically, we expect that $V_{min}$ provides the dominant potential energy
for the inflationary evolution of the survived light K\"ahler moduli. The COBE
normalization of density perturbations at the Hubble exit $\delta_{H}\approx 1.92
\times 10^{-5}$ then demands\( \left({V_{min}}/{\epsilon_{1}}\right)^{\frac{1}{4}}
\sim 0.004 \sim 5.4\times 10^{16}~\textrm{GeV} \) at the Hubble exit, where $\epsilon_{1}$
is the first slow-roll parameter. The range of $\epsilon_{1}$ at the horizon exit is
$\epsilon_{1}\sim 10^{-12}$ for typical values of the microscopic parameters. Thus
the inflationary energy scale should be rather low in such models, $V_{min}\sim
10^{-19}$\cite{quevedo062}. This requirement can be easily satisfied by making some
fine-tuning on the involved stringy parameters and, in fact, there exists a landscape
in parameter space in favor of phenomenology and cosmology. One possibility that
we will focus on in the next section is given by an orientifold model of $h^{(1,1)}=4,
~h^{(1,2)}\approx 100$, with geometrical parameters $\alpha_0 = {1}/{9\sqrt 2}$ and
$\lambda_2 =\lambda_3 =\lambda_4 =1$\cite{bond}. The complex structure moduli including
dilaton are supposed to be fixed by imaginary self-dual 3-form fluxes in such
a manner that after these moduli are stabilized we have $W_0=100, ~g_s\approx 0.132$
and consequently $\xi\approx 10$. The other microscopic parameters in the model are
designated as,
\[
\begin{array}{ll}
& A_2=1, ~~A_3 =A_4 \approx 1.075\times 10^{-4}, ~~a_2=a_3=a_4={2\pi}/{300}, ~~\\
& \gamma \approx 1.026\times 10^{-5}.
\end{array}
\]
For such a model, numerical calculation tells us that ${\mathcal V}_f \approx 2.8\times
10^8$\footnote{That is, ${\mathcal V}_f \approx 2.8\times 10^8~l_{s}^{6}$, where
$l_{s}= 2 \pi\sqrt{\alpha^{\prime}}$.} and $V_{min}\approx 4.035\times
10^{-19}$\footnote{Correspondingly, the density of the vacuum energy during inflation
is of the scale $M \approx V^{\frac{1} {4}}_{min} \approx 3.075 \times 10^{14}~
\textrm{GeV}$. This setup is consistent with the upper bound $V^{\frac{1}{4}}< 3.8
\times 10^{16}~\textrm{GeV}$ from the present WMAP dataset\cite{riotto}.}. The volume
modulus $\tau_1$ and heavy modulus $\tau_2$ are found to be frozen at $\tau_{1,f} \approx
4.28\times 10^5$ and $a_2\tau_{2,f}\approx 13.81 \approx 0.71 \ln{\mathcal V}_f$
respectively, as expected.

The large volume approach to heavy K\"ahler moduli stabilization naturally provides
a platform for having a stringy inflation scenario, where the survived light
K\"ahler moduli, say $\tau_i$ for $i=k+1, \cdots, n$ ($k\geq 2$) play the roles of
the multifield inflation drivers. The effective potential for these light moduli is
\begin{equation}\label{eq: 12}
%\begin{array}{lll}
V \approx  V_{min} + \sum\limits_{i=k+1}^n \frac{8(a_i A_i)^2 \sqrt{\tau_i}}
{3{\mathcal V}_f\lambda_i \alpha_0} e^{-2a_i\tau_i}
+ \sum\limits_{i=k+1}^n \frac{4a_iA_iW_0\tau_i}{{\mathcal V}_f^2}
e^{-a_i\tau_i} \cos(a_i\theta_i)~~~~~~
%\end{array}
\end{equation}
and these scalar fields are assumed to be initially far from the minimum of potential
$V$ (Please distinguish $V_{min}$ from the minimum of $V$. The latter is expected to
be zero). Because in Eq.(\ref{eq: 12}) $V_{min}$ is one order of the fixed Calabi-Yau
volume in the magnitude greater than the other scalar field dependent terms, the fixed
volume ${\mathcal V}_f$ and all of the stabilized heavy K\"ahler moduli as well as the
associated axions will remain frozen during inflation\footnote{The unique exception is
the axion associated with the volume modulus, $\theta_1$. It is free of stabilization
in the large volume approach. However, this is harmless.}. These frozen heavy moduli
will not create the harmful contributions to the masses of the light scalar fields.

The form of potential (\ref{eq: 12}) implies that there is a stringy landscape to realize
the cosmological inflation in IIB flux compactification at the large volume limit. If we
consider the type IIB compactification on the Calabi-Yau orientifolds with $h^{(1,1)}=n=3$,
for example, we will get K\"ahler modulus inflation\cite{quevedo062} when the axion
$\theta_3$ is stabilized at the potential minimum or roulette inflation\cite{bond}
when $\theta_3$ is dynamical. In this paper, we consider the IIB compactification on
a Calabi-Yau orientifold of $h^{(1,1)}=n \geq 4$, with almost all K\"ahler moduli
(including ${\mathcal V}$, $\tau_1$ and $T_i=\tau_i + i \theta_i$ for $i=2, 3, \cdots,
n-2$) have been stabilized. The only exception is for the two light moduli $T_{n-1}
= \tau_{n-1} + i \theta_{n-1}$ and $T_n=\tau_n + i \theta_n$ which correspond to four
scalar fields. The dynamics of this four-field system is governed by an effective potential
\begin{equation}\label{eq: 13}
\begin{array}{lll}
V & \approx & V_{min} + \sum\limits_{i=n-1}^n \frac{8(a_i A_i)^2 \sqrt{\tau_i}}
{3{\mathcal V}_f\lambda_i \alpha_0} e^{-2a_i\tau_i}\\
&    & + \sum\limits_{i=n-1}^n \frac{4a_iA_iW_0\tau_i}{{\mathcal V}_f^2}
e^{-a_i\tau_i} \cos(a_i\theta_i)
\end{array}
\end{equation}
Recall that the effective superpotential for these fields is a sum of two exponential
terms, the potential (\ref{eq: 13}) might be a large volume version of the better
racetrack inflation model\cite{quevedo061}.

The detailed investigation to this refined better racetrack model is scheduled to be
published in the near future\cite{yang}. In the present paper, we do what Conlon and
Quevedo have done by fixing the axions to the potential minimum and ignoring the double
exponentials, but regard both light K\"ahler moduli $\tau_3$ and $\tau_4$ as the relevant
scalar fields (Here we take $n=4$ for simplicity). This will yield an assisted-like
extension of the single field K\"ahler modulus inflation model. The potential energy
for such a two-field system reads,
\begin{equation}\label{eq: 14}
V \approx V_{min}
- \frac{4 a_3 A_3W_0\tau_{3}}{{\mathcal V}_f^2} e^{-a_3\tau_3}
- \frac{4 a_4 A_4W_0\tau_{4}}{{\mathcal V}_f^2} e^{-a_4\tau_4}
\end{equation}
The kinetic Lagrangian is given by ${\mathscr L}_{K} = -\sum_{i,j=3}^4 K_{i\bar{j}}
\partial_\mu \tau_i \partial^\mu \tau_j$. Explicitly,
\[
{\mathscr L}_{K}
=  - \frac{3\alpha_0 \lambda_3}{8{\mathcal V}_f\sqrt{\tau_3}}
(\partial \tau_3)^2 - \frac{3\alpha_0 \lambda_4}{8{\mathcal V}_f
\sqrt{\tau_4}} (\partial \tau_4)^2
- \frac{9\alpha^2_0 \lambda_3 \lambda_4 \sqrt{\tau_3
\tau_4}}{4{\mathcal V}_f^2} (\partial_\mu \tau_3) (\partial^\mu \tau_4)
\]
Let
\begin{equation}\label{eq: 15}
\varphi_i = \sqrt{\frac{4\alpha_0 \lambda_{i+2}}{3{\mathcal V}_f}}\tau_{i+2}^{3/4}
\end{equation}
with $i=1,~2$, we get,
\begin{equation}\label{eq: 16}
{\mathscr L}_{K} = -\frac{1}{2}\partial^\mu \varphi_1\partial_\mu \varphi_1
-\frac{1}{2}\partial^\mu \varphi_2 \partial_\mu \varphi_2
- \frac{9}{4}\varphi_1 \varphi_2
\partial^\mu \varphi_1 \partial_\mu \varphi_2
\end{equation}
The potential (\ref{eq: 14}) can be recast as
\begin{equation}\label{eq: 17}
V(\varphi_1, \varphi_2)  = V_{min} + V_1\psi(\varphi_1)e^{-\beta_1 \psi(\varphi_1)}
+ V_2\psi(\varphi_2)e^{-\beta_2 \psi(\varphi_2)}
\end{equation}
in terms of the newly defined fields. In Eq.(\ref{eq: 17}) we have defined $\psi(\varphi_i)
= \varphi_i^{4/3}$, $\beta_i=a_{i+2}(\frac{3{\mathcal V}_f}{4\alpha_0
\lambda_{i+2}} )^{\frac{2}{3}}$ and $V_i = - \frac{4\beta_i A_{i+2} W_0}{{\mathcal V}_f^2}$
for convenience.

\section{Two-field inflation model with non-canonical lagrangian}\label{sec:model}

In this section, we intend to study the probable implications of the above two-field model
on cosmology. The model emerges from the large volume approach to the moduli stabilization
of Type IIB superstring flux compactification and is described by the effective action
\begin{eqnarray}
& S = & \int d^4 x \sqrt{-G} \Big[ \frac{1}{2}{\mathcal R}
-\frac{1}{2}\partial^\mu \varphi_1\partial_\mu \varphi_1
-\frac{1}{2}\partial^\mu \varphi_2 \partial_\mu \varphi_2 \nonumber \\
&  &~~~~~~~~~~~~~~~~~~~~~~ - \frac{9}{8} \varphi_1 \varphi_2 \partial^\mu \varphi_1 \partial_\mu \varphi_2 -V(\varphi_1, \varphi_2) \Big ] \label{eq: 18}
\end{eqnarray}
with $V(\varphi_1, \varphi_2)$ given in Eq.(\ref{eq: 17}). Similar to the assisted
inflation model\cite{lms}, the potential $V(\varphi_1, \varphi_2)$ consists of
the sum of two exponentials of the scalar fields. The model can be thought of
as a simplified version of the large volume better racetrack model in which two axionic
fields have been stabilized at the potential minimum. In action (\ref{eq: 18}), $G$
stands for the determinant of 4-dimensional metric tensor $G_{\mu\nu}$, whose Ricci
scalar curvature is denoted by ${\mathcal R}$.  From the assigned stringy parameters
for the model, the involved phenomenological parameters take values of $V_{min}\approx
4.035 \times 10^{-19}, ~V_{1} = V_2 \approx -2.212\times 10^{-14}$ and $\beta_1 = \beta_2
\approx 40337.578$ respectively. The potential has a Minkowski minimum, $V(\varphi_1,
\varphi_2)\approx 0$, at $\varphi_{1} = \varphi_{2} \approx 3.51 \times
10^{-4}$\footnote{Correspondingly, the light K\"ahler moduli at this Minkowski minimum
take values $\tau_{3}=\tau_{4}\approx 47.75$.}, however, it is very flat far away from
this minimum.
\begin{figure}[ht]
\begin{center}
\includegraphics{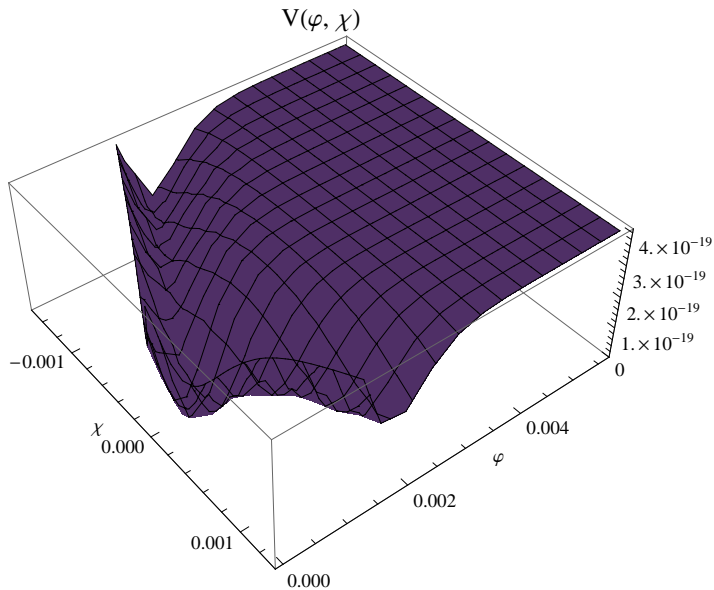}
\caption{\small{The potential surface of our two-field model defined in Eqs. (\ref{eq: 20})
and (\ref{eq: 21}) where the values $V_{min}\approx 4.035 \times 10^{-19}, ~V_{1} = V_2
\approx -2.212\times 10^{-14}$ and $\beta_1 = \beta_2 \approx 40337.578$ have been assigned
to the parameters. The potential has a Minkowski minimum at $\varphi \approx 4.97\times
10^{-4}$ and $\chi =0$ and is very flat along the $\varphi$-direction far away from this
minimum.} }
\label{potential}
\end{center}
\end{figure}

The crossing term in the kinetic lagrangian brings on much unnecessary inconvenience.
To overcome it, we redefine the independent scalar fields by
\begin{equation}\label{eq: 19}
\varphi_1 = (\varphi + \chi)/{\sqrt{2}}, ~~~
\varphi_2 = (\varphi - \chi)/{\sqrt{2}}. ~
\end{equation}
The action becomes
\begin{eqnarray}
& S = & \int d^4 x \sqrt{-G} \bigg[ \frac{1}{2}{\mathcal R}
-\cos^2 \alpha ~\partial^\mu \varphi \partial_\mu \varphi
-\sin^2 \alpha ~\partial^\mu \chi \partial_\mu \chi
- V(\varphi, \chi) \bigg ] \nonumber  \\
&   & \label{eq: 20}
\end{eqnarray}
where
\begin{equation}\label{eq: 21}
\begin{array}{lll}
V(\varphi, \chi) & = ~ V_{min} & + ~V_1\psi((\varphi + \chi)/\sqrt2)
e^{-\beta_1 \psi((\varphi + \chi)/\sqrt2)} \\
&  & + ~V_2\psi((\varphi - \chi)/\sqrt2)e^{-\beta_2 \psi((\varphi - \chi)/\sqrt2)}
\end{array}
\end{equation}
and $\cos2 \alpha =\frac{9}{8}(\varphi^2 - \chi^2)$. Because in large volume limit
the magnitudes of both $\varphi$ and $\chi$ could be much smaller than \emph{one},
the introduction of the field-dependent auxiliary quantity $\alpha(\varphi, \chi)$
by its cosine value is reasonable. The appearance of $\alpha(\varphi, \chi)$ in
the kinetic lagrangian implies that we get a nontrivial diagonal metric in field
space, which does not coincide with the known non-canonical two-field kinetic Lagrangian
studied in \cite{langlois}, and is expected to bring some new features in cosmological
application. In terms of newly defined scalars, the Minkowski minimum of the potential
(\ref{eq: 21}) occurs at $\varphi \approx 4.97\times 10^{-4}$ and $\chi =0$. Departure
from this minimum is a plateau mainly along the $\varphi$-direction which might be
adequate for inflation.

We begin with the equations of motion of the homogeneous background fields. Consider a
spatially flat Robertson-Walker spacetime
\begin{equation}\label{eq: 22}
ds^2_4 = -dt^2 + a(t)^2 d{\bf x}^2
\end{equation}
Here $t$ is the cosmic time. It is much convenient to formulate the equations of motion of
the scalar fields in terms of the efold time $n=\ln\left(a(t)/a_{ini}\right)$, by which
the equations of motion of the scalar fields are decoupled from the metric evolution.
For the sake of convenience, we define a homogeneous inflaton field $\sigma(t)$ through
its velocity,
\begin{equation}\label{eq: 23}
\dot{\sigma}=\sqrt{2\cos^2 \alpha \dot{\varphi}^2 + 2\sin^2 \alpha \dot{\chi}^2}
\end{equation}
In Eq.(\ref{eq: 23}) and hereinafter, a dot stands for a derivative with respect to
the efold time $n$. The equations of motion of the scale factor in metric (\ref{eq: 22})
and the homogeneous background scalar fields are found to be,
\begin{eqnarray}
&  & H^2 = \frac{V}{3-\dot{\sigma}^2/2} \label{eq: 24}\\
&  & \frac{\dot{H}}{H} = - \frac{1}{2}\dot{\sigma}^2  \label{eq: 25}\\
&  & \frac{\ddot{\varphi}
-\tan \alpha \partial_{\varphi} \alpha (\dot{\varphi}^2 + \dot{\chi}^2)
-2 \tan \alpha \partial_{\chi}\alpha \dot{\varphi}\dot{\chi}}
{3 -\dot{\sigma}^2/2}+\dot{\varphi} = - \frac{\sec^2 \alpha
\partial_{\varphi}V}{2V} \label{eq: 26}\\
&  & \frac{\ddot{\chi}
+\cot \alpha \partial_{\chi} \alpha (\dot{\varphi}^2 + \dot{\chi}^2)
+2 \cot \alpha \partial_{\varphi}\alpha \dot{\varphi}\dot{\chi}}
{3 -\dot{\sigma}^2/2}+\dot{\chi} = - \frac{\csc^2 \alpha
\partial_{\chi}V}{2V} \label{eq: 27}
\end{eqnarray}
where $V$ is the abbreviation of the potential $V(\varphi,\chi)$, $H=da/adt$ is the Hubble
parameter but $\dot{H}=dH/dn$. Inflation occurs for $\frac{d^2a}{dt^2}>0$. Hence, having
an inflation driven by the above two scalar fields requires
\begin{equation}\label{eq: 28}
\epsilon_1 = - \frac{\dot{H}}{H} = \dot{\sigma}^2/2 < 1
\end{equation}
$\epsilon_1$ is the so-called first slow-roll parameter\footnote{The slow-roll
approximation is in general figured in terms of numerous Hubble flow
functions\cite{ringeval}, among which the first two are $\epsilon_1 = -\frac{\dot{H}}{H}$
and $\epsilon_2 = \frac{\dot{\epsilon_1}}{\epsilon_1}$. }. Is there an inflationary epoch
with $\epsilon_1 < 1$ for our two-field model? To have a definite answer to this question,
we have to numerically solve Eqs.(\ref{eq: 24}-\ref{eq: 27}) under appropriate initial
conditions. In fact, it is sufficient to integrate Eqs.(\ref{eq: 26}) and (\ref{eq: 27})
only, since these scalar field equations are decoupled from the metric evolution.

From the Cauchy theorem, the solution to Eqs.(\ref{eq: 26}) and (\ref{eq: 27}) is unique
provided the initial fields and initial field velocities are given at some initial instant
$n=n_{ini}$. However, just as pointed out by Ringeval\cite{ringeval}, the attractor
behavior induced by the friction terms erase any effect associated with the initial field
velocities after a few efolds. The integration of Eqs.(\ref{eq: 26}) and (\ref{eq: 27})
depends essentially upon the initial field values only. The attractor behavior does also
ensure the stability of forward numerical integration schemes, so we use a Runge-Kutta
integration method of order four. Numerically, we take $n_{ini} \approx
-4.60516$\footnote{See Eq.(\ref{eq: 47}) below for explanation.} and $\varphi_{ini} \approx
5.67\times 10^{-3}$. It follows from Eq.(\ref{eq: 19}) that, if $\chi_{ini}$
vanished, $\varphi_{1, ini}=\varphi_{2, ini}$, the evolutions of $\varphi_1$ and
$\varphi_2$ would exactly be the same, our multifield model would effectively reduce into
a single field model. To have a small but significant deviation from such an indifferent
situation, we take $\chi_{ini} \approx -1.13\times 10^{-5}$. Such a choice for initial
fields ensures them far away from their possible values at the Minkowski minimum, and
correspondingly the light K\"ahler moduli are initially set to be $\tau_{3}\approx 1224.14$
and $\tau_{4}\approx 1230.67$, respectively. The initial field velocities are chosen on
the attractor by setting
\begin{equation}\label{eq: 29}
\dot{\varphi}_{ini} = - \partial_{\varphi}\ln V|_{\varphi_{ini}, ~\chi_{ini}}, ~~~~
\dot{\chi}_{ini} = - \partial_{\chi}\ln V|_{\varphi_{ini}, ~\chi_{ini}}.
\end{equation}
\begin{figure}[ht]
\begin{center}
\includegraphics{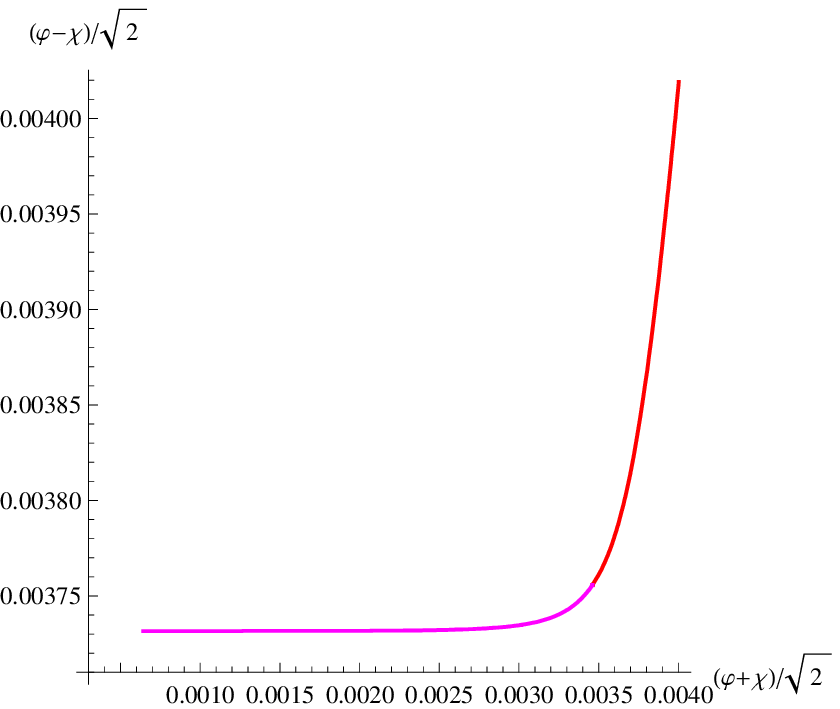}\hfill
\includegraphics{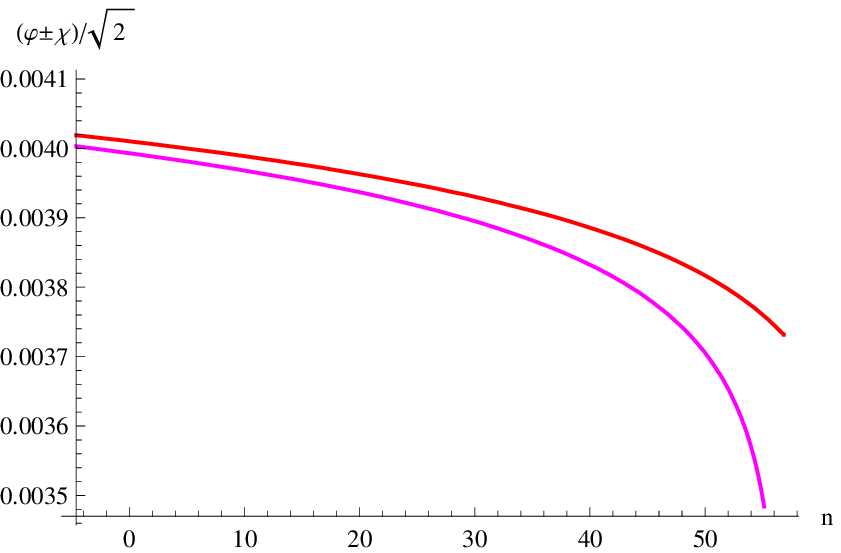}
\caption{\small Classical inflationary trajectory of the two-field model under
consideration, for efolds $-4.60516\leq n \leq 56.83847$. The inflationary epoch
starts at $n\approx -4.60516$ where we put the initial conditions $\varphi_{ini}
\approx 5.67\times 10^{-3}$ and $\chi_{ini} \approx-1.13\times 10^{-5}$.
In the first $35$ efolds the trajectory is approximately along the $(\varphi -
\chi)$-direction (red curve). When $n \gtrsim 30$, the fields $(\varphi \pm \chi)/\sqrt2$
begin to lose their synchrony in evolution. A steep turn in field space takes place
roughly at $n\approx 55.85$ where $\varphi\approx 5.04\times 10^{-3}$,
$\chi \approx -2.57\times 10^{-4}$ and after that $(\varphi - \chi)\sqrt2$ does no
longer evolve but $(\varphi + \chi)/\sqrt2$ continues to decrease. During the last efold,
$55.85\leq n \leq 56.83847$, the trajectory is along the $(\varphi + \chi)$-direction
(magenta curve). The inflation approaches to its end at $n\approx 56.83847$, where
$\varphi\approx 3.34\times 10^{-3}$, $\chi \approx -1.94\times 10^{-3}$ and $\epsilon_{1}
\approx 0.97$.}
\label{trajectory}
\end{center}
\end{figure}
%\vskip -14pt
Figure \ref{trajectory} gives our numerical solution to the classical trajectory of
the scalar fields for $-4.60516\leq n \leq 56.83847$. During most of the efolds,
$-4.60516\leq n \leq 55.85$, the scalars roll very slowly and the trajectory is
almost in the $(\varphi - \chi)$-direction. Being a two-field model means that
the trajectory is not a straight line and in fact a sharp turn on trajectory occurs
roughly at $n\approx 55.85$. Within the last efold time interval the trajectory becomes
another straight line along the $(\varphi + \chi)$-direction. Plugging this solution
for $\varphi(n)$ and $\chi(n)$ into Eqs.(\ref{eq: 24}) and (\ref{eq: 25}) determines
the evolution of Hubble parameter and the first slow-roll parameter $\epsilon_1$.
Our choice of the initial conditions ensures both $\epsilon_1$ and $\epsilon_2$ much
smaller than \emph{one} for most of the efolds. In particular, $\epsilon_1 \approx
4.82\times 10^{-12}$ and $\epsilon_2\approx 0.036$ at the Hubble exit ($n \approx
2$). There is indeed an inflationary epoch in our model during which the Hubble parameter
hardly changes. The potential $V$ is very flat along the classical trajectory and,
just as in the case of single field case\cite{quevedo062}, it is almost a constant,
$V \approx 4.035 \times 10^{-19}$, for most of inflation. Only within the final fractions
of the last efold does a steep decay of potential take place. When $n$ impends over $56.83847$,
the potential sharply drops off from the preceding constant to a metastable minimum
with $V\approx 2.725\times 10^{-19}$, where $\epsilon_1\approx 0.97$ and the inflation
is close to its end.
\begin{figure}[ht]
\begin{center}
\includegraphics{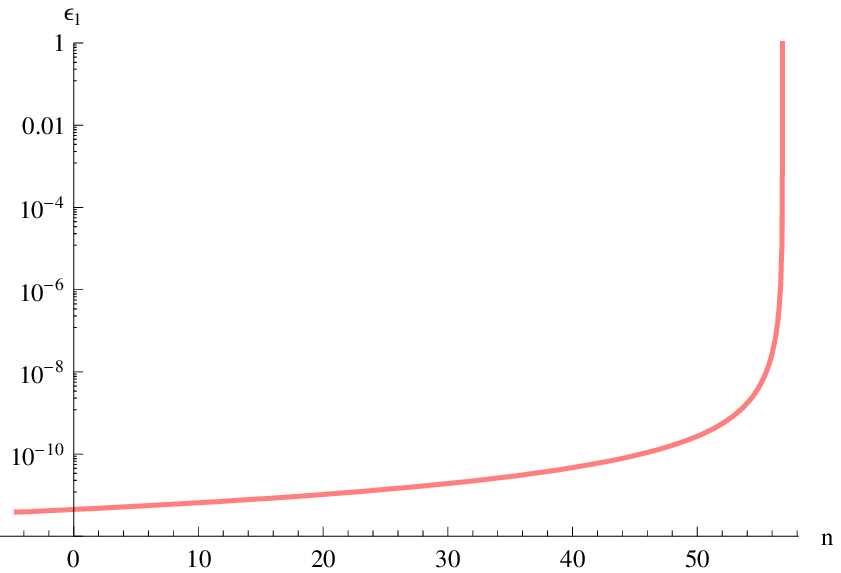}
\caption{\small The first slow-roll parameter $\epsilon_1$ during the efold time interval
$-4.60516\leq n \leq 56.83847$ for our two-field model. $\epsilon_1$ is much smaller than
\emph{one} for most of the efolds, and particularly $\epsilon_1\approx 4.82 \times
10^{-12}$ at $n = 2$, implying that the potential is almost flat during the epoch.
$\epsilon_1$ increases drastically during the last several efolds. When $n=56.83847$,
$\epsilon_1 \approx 0.97$, the inflation tends to end. }
\label{fig23}
\end{center}
\end{figure}
%\vskip -14pt

The apodictic deviation of trajectory from an entire straight line shows that what we
consider is a genuine multifield model. Provided that the isocurvature perturbation
decays fast enough on the super-Hubble scales, it will not influence the super-Hubble
curvature perturbation heavily even if the trajectory in field space deviates deadly
from the straight-line one of a single field model. The steep decay of the isocurvature
perturbations takes place in many supergravity or string theory inspired multifield
models, and in particular in the string theory inspired roulette model\cite{bond}.
The inflation in these models is effectively driven by a single scalar field where one
can use the single field approximation to account for the super-horizon power spectrum
of the curvature perturbation. However, the same is not true for the present situation.
In our two-field model, the decay of the isocurvature perturbation on super-Hubble scales
is not fast and the spectrum of the curvature perturbation in the vicinity of the Hubble
crossing is remarkably deviated by that at the end of inflation, where the correlation
between the curvature and isocurvature modes becomes sufficiently strong so that the final
curvature perturbation is heavily sourced by the isocurvature perturbation.

We are going to confirm the above conclusion for the two-field model in consideration
through numerical calculations of the power spectra of the linear perturbations and
their correlation. Due to the fact that at the Hubble-exit $\epsilon_1\approx 4.82\times
10^{-12}$, there is no important tensor fluctuations in our model. This might be a common
feature of the stringy inflation models where the inflation is driven by some closed
string moduli\cite{quevedo04, quevedo061, quevedo062, bond}. So we consider only the
scalar perturbations. Because the matter is composed of two scalar fields, the stress
tensor is diagonal and the perturbed metric is of the form
\begin{equation}\label{eq: 30}
ds^2_4 = -\left[1 + 2\Psi(t, {\bf x})\right]dt^2
+ a(t)^2 \left[1 -2 \Psi(t,{\bf x})\right]d{\bf x}^2
\end{equation}
in the longitudinal gauge. The scalar fields are decomposed into their homogeneous
backgrounds plus the anisotropic perturbations:
\begin{equation}\label{eq: 31}
\varphi(t,{\bf x})=\varphi(t) + \delta \varphi(t, {\bf x}), ~~~~
\chi(t,{\bf x})=\chi(t) + \delta \chi(t, {\bf x}).
\end{equation}
To facilitate solving the evolution equations of these cosmological perturbations, we can
alternatively decompose them into the instantaneous curvature and isocurvature components,
relying on the fact that the isocurvature modes do only source the curvature perturbations
approximately\cite{marco, gordon}. The curvature and isocurvature perturbations are
parallel with and orthogonal to the trajectory of the homogeneous inflaton field
$\sigma(t)$, respectively. To define them we first introduce a time dependent angle $\theta$,
\begin{equation}\label{eq: 32}
\cos\theta = \sqrt{2}\cos \alpha \frac{\dot{\varphi}}{\dot{\sigma}},~~~
\sin\theta = \sqrt{2}\sin \alpha \frac{ \dot{\chi}}{\dot{\sigma}}.
\end{equation}
with which we can recast the background equations (\ref{eq: 26}) and (\ref{eq: 27}) as
\begin{eqnarray}
&  & \ddot{\sigma} + (3- \dot{\sigma}^2/2)\dot{\sigma} + \frac{V_\sigma}{H^2} = 0
\label{eq: 33} \\
&  & \dot{\theta} + \frac{V_s}{H^2 \dot{\sigma}} + \frac{\dot{\sigma}}{\sqrt2}
(\cos\theta \sec \alpha \alpha_\chi + \sin\theta \csc \alpha \alpha_\varphi) = 0
\label{eq: 34}
\end{eqnarray}
The curvature perturbation $\delta \sigma(t,{\bf x})$ and isocurvature perturbation
$\delta s(t,{\bf x})$ are defined through a linear transformation in field space
\begin{equation}\label{eq: 35}
\left[
\begin{array}{r}
\delta \sigma(t,{\bf x}) \\ \delta s(t,{\bf x})
\end{array}
\right]
= \sqrt2
\left[
\begin{array}{rr}
\cos \alpha \cos \theta  & \sin \alpha \sin \theta \\
-\cos \alpha \sin \theta & \sin \alpha \cos \theta
\end{array}
\right]
\left[
\begin{array}{r}
\delta \varphi(t,{\bf x}) \\ \delta \chi(t,{\bf x})
\end{array}
\right]
\end{equation}
From Eq.(\ref{eq: 35}) we have $\delta s={\sin2\alpha}(\dot{\varphi}\delta \chi - \dot{\chi}
\delta \varphi)/{\dot{\sigma}}$. The isocurvature perturbations turn out to be the entropy
perturbations.

It is worthwhile to stress that, among the linear scalar perturbations $\Psi(t,{\bf x}),
\delta \sigma(t,{\bf x})$ and $\delta s(t,{\bf x})$, only two of them are independent.
The independent scalar perturbations can also be chosen as the gauge invariant
Mukhanov-Sasaki variable
\begin{equation}\label{eq: 36}
Q_{\sigma}(t,{\bf x}) = \delta \sigma(t,{\bf x}) + \dot{\sigma}\Psi(t,{\bf x})
\end{equation}
and $\delta s(t,{\bf x})$. In terms of these gauge invariant perturbations, the perturbed
Klein-Gordon equations and perturbed Einstein equations\cite{ringeval} can be unified into
two coupled second order differential equations:
\begin{eqnarray}
&  & \ddot{Q}_{\sigma}+(3-\dot{\sigma}^2/2)\dot{Q}_{\sigma} + \frac{2V_s}{H^2 \dot{\sigma}}
\dot{\delta s} - \frac{1}{H^2 a^2}\nabla^2 Q_{\sigma}
+ \frac{{\mathcal C}_{\sigma \sigma}}{H^2}Q_{\sigma}
+ \frac{{\mathcal C}_{\sigma s}}{H^2}\delta s = 0 \label{eq: 37} \\
&  & \ddot{\delta s} + (3 - \dot{\sigma}^2/2)\dot{\delta s} - \frac{2V_s}{H^2 \dot{\sigma}}
\dot{Q}_{\sigma} - \frac{1}{H^2 a^2}\nabla^2\delta s
+ \frac{{\mathcal C}_{s\sigma}}{H^2}Q_{\sigma}
+ \frac{{\mathcal C}_{ss}}{H^2}\delta s = 0 \label{eq: 38}
\end{eqnarray}
where
\begin{eqnarray}
&  {\mathcal C}_{\sigma\sigma} = & V_{\sigma \sigma}
+ H^2(3-\dot{\sigma}^2/2)\dot{\sigma}^2
- \frac{1}{H^2}\left(\frac{V_s}{\dot{\sigma}}\right)^2
+ 2\dot{\sigma}V_{\sigma} \nonumber \\
&  & ~~~~~~ + \frac{\sqrt2 V_{\sigma}}{\sin^2 2\alpha}(\cos2\theta
-\cos2\alpha)(\cos\theta \sin \alpha \alpha_{\varphi}
+ \sin \theta \cos \alpha \alpha_{\chi}) \nonumber \\
&  & ~~~~~~ + \frac{\sqrt2 V_s}{\sin^2 2\alpha}
\Big[ \cos\theta \cos \alpha (\cos 2\theta + \cos 2\alpha -2) \alpha_{\chi} \nonumber \\
&  & ~~~~~~ ~~~~~~~~~~~~~~~~~~~~~~~
- \sin\theta \sin \alpha (\cos 2\theta + \cos 2\alpha + 2) \alpha_{\varphi} \Big] \label{eq: 39}\\
& {\mathcal C}_{\sigma s} = & (6 + \dot{\sigma}^2)\frac{V_s}{\dot{\sigma}}
+ \frac{2V_\sigma V_s}{H^2 \dot{\sigma}^2} + 2V_{\sigma s} \nonumber \\
&  & ~~~~~~ + \frac{2\sqrt2 V_\sigma }{\sin^2 2\alpha}(\cos2\theta
- \cos 2\alpha) (\cos\theta \cos \alpha \alpha_\chi
- \sin\theta \sin \alpha \alpha_\varphi) \nonumber \\
&  & ~~~~~~ - \frac{2\sqrt2 V_s}{\sin^2 2\alpha} (\cos 2\theta
+ \cos 2\alpha ) (\cos\theta \sin \alpha \alpha_\varphi
+ \sin\theta \cos \alpha \alpha_\chi) \label{eq: 40} \\
& {\mathcal C}_{s\sigma} = & -6\frac{V_s}{\dot{\sigma}}
- \frac{2V_\sigma V_s}{H^2 \dot{\sigma}^2}
+ V_s\dot{\sigma} \label{eq: 41} \\
&  {\mathcal C}_{ss} = & V_{ss} - \frac{1}{H^2}\left(\frac{V_s}{\dot{\sigma}}\right)^2
- \frac{\dot{\sigma}^2}{\sin 2\alpha} (\alpha_{\varphi \varphi}
- \alpha_{\chi \chi}) \nonumber \\
&  & ~~~~~~ - \frac{\sqrt2 V_s}{\sin^2 2\alpha}(\cos2\theta
+ \cos2\alpha)(\cos\theta \sin \alpha \alpha_{\chi}
- \sin \theta \cos \alpha \alpha_{\varphi}) \nonumber \\
&  & ~~~~~~ + \frac{\sqrt2 V_\sigma}{\sin^2 2\alpha}
\Big[ \cos\theta \sin \alpha (2 -\cos 2\theta + \cos 2\alpha ) \alpha_{\varphi} \nonumber \\
&  & ~~~~~~ ~~~~~~~~~~~~~~~~~~~~~~~
- \sin\theta \cos \alpha (2 + \cos 2\theta - \cos 2\alpha ) \alpha_{\chi} \Big] \label{eq: 42}
\end{eqnarray}
As usually done, we will work with the Fourier components of the perturbations, $Q_{\bf
k}(n)$ and $\delta s_{\bf k}(n)$, with $k$ a given comoving wave number. For sake of
convenience, in Eqs.(\ref{eq: 37}-\ref{eq: 42}) we have defined various derivatives of
the potential with respect to the curvature and isocurvature directions, which are
associated with the derivatives of the potential with respect to the original scalars
$\varphi$ and $\chi$ through the linear transformations
\begin{equation}\label{eq: 43}
\left[ \begin{array}{c} V_\sigma \\ V_s \end{array} \right]
= {\mathcal T} \left[\begin{array}{c} V_\varphi  \\ V_\chi  \end{array} \right]
\end{equation}
and
\begin{equation}\label{eq: 44}
\left[ \begin{array}{cc} V_{\sigma \sigma} & V_{\sigma s} \\ V_{\sigma s} & V_{ss}
\end{array} \right]
= {\mathcal T}
\left[\begin{array}{cc}
V_{\varphi \varphi}  & V_{\varphi \chi} \\
V_{\varphi \chi}  & V_{\chi \chi}
\end{array} \right] {\mathcal T}^T
\end{equation}
The transformation matrix ${\mathcal T}$ reads,
\begin{equation}\label{eq: 45}
{\mathcal T} = \frac{1}{\sqrt2}
\left[ \begin{array}{rr} \cos\theta \sec \alpha & \sin\theta \csc \alpha \\
-\sin\theta \sec \alpha & \cos\theta \csc \alpha \end{array} \right]
\end{equation}

To study the generation of perturbations from the vacuum fluctuations, we have to solve
the closed set of Eqs.(\ref{eq: 37}) and (\ref{eq: 38}) under appropriate initial conditions.
Though the curvature and isocurvature fluctuations are finally coupled to each other
beyond the Hubble crossing, they might be statistically independent deeply inside the
Hubble radius. The initial efold time $n_{ini}$ at which the perturbations are thought of
as mutually independent must be much less than the Hubble crossing instant $n_*$ for
Fourier mode perturbations of wave number $k$, $k\lesssim a(n_*)H(n_*)$. The value of $n_*$
depends on the efold number during which the universe reheated before the starting of the
radiation era\cite{ringeval} and it is known\cite{liddle} that typically $40\lesssim N_*
\lesssim 60$ for $N_* =n_{end}-n_*$. Without knowledge of details of the reheating mechanism
we can not be more precise to say what $n_*$ should be. However, if $n_*$ is known somehow,
the efolding number $N_*$ of inflation after horizon crossing can be estimated from the
assumption that the reheat temperature is of the same order as but slightly lower than
the energy scale of inflation, $T_{reh}\approx 3 \times 10^{14}$ GeV, through the
formula\cite{quevedo061},
\begin{equation}\label{eq: 46}
N_* \approx 53 + \ln\left(\frac{T_{reh}}{10^{13} ~\textrm{GeV}}\right) \lesssim 57~.
\end{equation}
We will simply set $n_* = 2$. Then $N_*\approx 54.83847$, compatible with the constraint
from cosmological observation. The pivot wave number of the Fourier mode perturbation is
taken as $k=a(0)H(0)$ for which the initial efold time $n_{ini}$ can be defined by a
cutoff equation\cite{salopek}
\begin{equation}\label{eq: 47}
\frac{k}{a(n_{ini})H(n_{ini})}=\frac{H(0)}{e^{n_{ini}}H(n_{ini})}=C_q
\end{equation}
The constant $C_q$ should be sufficient large in order that the probable interferences
between the initial curvature and isocurvature modes remain negligible\cite{niemeyer}.
We take $C_q \approx 100$ for concreteness, which leads to $n_{ini}=-4.60516$. To take
into account the statistical independence of the initial perturbations in our numerical
procedure, we integrate Eqs.(\ref{eq: 37}) and (\ref{eq: 38}) twice: first with
the initial conditions,
\begin{equation}\label{eq: 48}
\begin{array}{l}
Q_{\sigma}(n_{ini})\approx Q_{ini}, ~~~\dot{Q}_{\sigma}(n_{ini})\approx -(1+i C_q)Q_{ini},\\
\delta s(n_{ini})\approx \dot{\delta s}(n_{ini})\approx 0~
\end{array}
\end{equation}
and second with the initial conditions,
\begin{equation}\label{eq: 49}
\begin{array}{l}
Q_{\sigma}(n_{ini})\approx \dot{Q}_{\sigma}(n_{ini})\approx 0, \\
\delta s(n_{ini})\approx \delta s_{ini},
~~~\dot{\delta s}(n_{ini})\approx -(1+i C_q)\delta s_{ini}.~~
\end{array}
\end{equation}
with $Q_{ini}=\delta s_{ini}={1}/{a(n_{ini})\sqrt{2k}}$. The curvature and isocurvature
perturbations are customarily described by ${\mathcal R}=Q_\sigma/\dot{\sigma}$ and
${\mathcal S}=\delta s/ \dot{\sigma}$, for which we get ${\mathcal R}_i$ and ${\mathcal
S}_i$ ($i=1, ~2$) after finishing two individual integrations of Eqs.(\ref{eq: 37}) and
(\ref{eq: 38}). ${\mathcal R}_{1}$ (${\mathcal S}_{2}$) can be thought of as the curvature
(isocurvature) perturbation without taking into account the coupling to the isocurvature
(curvature) perturbation, and ${\mathcal R}_{2}$ (${\mathcal S}_{1}$) does just come from
such a coupling. The final power spectra and their correlation are calculated as
follows\cite{tsujikawa}:
\begin{eqnarray}\label{eq: 50}
&  & {\mathscr P}_{\mathcal R} = \frac{k^3}{2\pi^2} \left(|{\mathcal R}_1|^2
+ |{\mathcal R}_2|^2 \right), ~~~
{\mathscr P}_{\mathcal S} = \frac{k^3}{2\pi^2} \left( |{\mathcal S}_1|^2
+ |{\mathcal S}_2|^2 \right) ,
~~~\nonumber \\
&  & {\mathscr C}_{\mathcal{RS}} = \frac{k^3}{2\pi^2} \left({\mathcal R}_1^\dagger {\mathcal S}_1
+ {\mathcal R}_2^\dagger {\mathcal S}_2\right) ~.
\end{eqnarray}
In fact, the correlation between the curvature and isocurvature perturbations can be simply
described by the so-called the relative correlation coefficient ${\mathscr C} =
{|{\mathscr C}_{\mathcal{RS}}|}/{\sqrt{{\mathscr P}_{\mathcal R}{\mathscr P}_{\mathcal S}}}$.
The value of ${\mathscr C}$ lies between $0$ and $1$, which measures to what extent the final
curvature perturbation results from the interactions with the isocurvature perturbation.

\begin{figure}[ht]
\begin{center}
\includegraphics{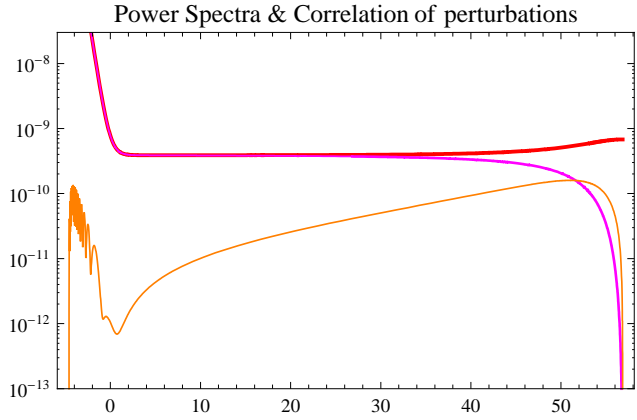}
\caption{\small Power spectra and correlation of perturbations. The red and magenta curves
represent respectively the spectra of curvature and isocurvature perturbations. The orange
curve describes the correlation between them. All these curves are plotted in the efold
time interval $-4.06516 \leq n \leq 56.83847$. }
\label{spectracorrelation}
\end{center}
\end{figure}
\begin{figure}[ht]
\begin{center}
\includegraphics{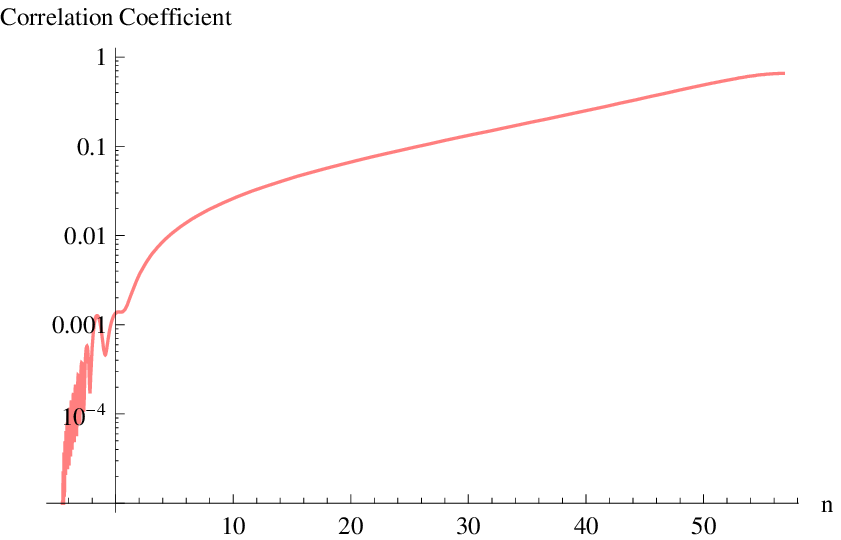}
\caption{\small Relative correlation coefficient ${\mathscr C}$ in the present model. It
vanishes initially, as expected. At $n=-1, ~0, ~1$ and $2$, ${\mathscr C}\approx 5.18
\times 10^{-4}, ~1.35\times 10^{-3}, ~1.69\times 10^{-3}$ and $3.6\times 10^{-3}$,
respectively. The coefficient increases monotonously on super-Hubble scales. When inflation
goes close to its end, $n\approx 56.83847$, ${\mathscr C} \approx 0.66$. }
\label{correlation}
\end{center}
\end{figure}

Our result of numerical integration for linear perturbations is displayed in Figure
\ref{spectracorrelation}, where the power spectra of curvature and isocurvature
perturbations as well as the correlation between them are plotted as functions of
the number $n$ of efolds. On super-Hubble scales the relative correlation coefficient
continues to increase, which is displayed in Figure \ref{correlation}.

\begin{figure}[ht]
\begin{center}
\includegraphics{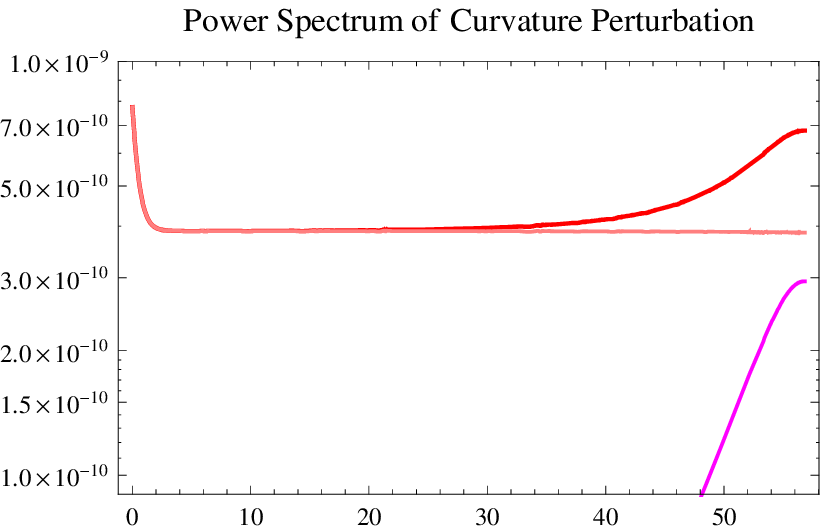}
\caption{\small Power spectrum of curvature perturbation in efold time interval $-4.60516
\lesssim n \lesssim 56.83847$. Red curve describes the evolution of the complete curvature
spectrum. Pink and magenta curves stand for the partial curvature spectra from
fluctuations along the inflationary trajectory and from the interactions with the
isocurvature modes, respectively. The curvature spectrum at the end of inflation is
${\mathscr P}_{\mathcal R}|_{n\approx 56.83847} \approx 6.79\times 10^{-10}$, which
is different from the curvature spectrum ${\mathscr P}_{\mathcal R}|_{n\approx 2} \approx
3.96\times 10^{-10}$ at Hubble-exit.}
\label{curvaturespectrum}
\end{center}
\end{figure}

The spectra drop off sharply on subhorizon scales until the supposed Hubble exit for the
Fourier mode perturbation is achieved. On superhorizon scales, however, the spectra
become sufficiently flat for $2 \lesssim n \lesssim 30$, with ${\mathscr P}_{\mathcal R}
\approx {\mathscr P}_{\mathcal S} \approx 3.96 \times 10^{-10}$ and correlator ${\mathscr
C}_{\mathcal{RS}}$ increases steadily from $1.46\times 10^{-12}$ to $5.041\times 10^{-11}$
($3.6\times 10^{-3} \lesssim {\mathscr C} \lesssim 0.132$). During the period of $n
\gtrsim 30$, though ${\mathscr C}_{\mathcal{RS}}$ increases at first but then drops off
steeply, ${\mathscr C}$ increases monotonously. For $n \gtrsim 30$, the spectrum of
curvature perturbation begins to increase while the spectrum of isocurvature
begins to decrease once more. The persistent increase of the curvature perturbation spectrum
for efold time interval $30\lesssim n \lesssim 56.6$ reflects a fact that the correlation
between the curvature and isocurvature perturbations during this time interval has become
sufficiently strong that it can no longer be neglected. The decay of isocurvature
perturbation spectrum accelerates in the final stage of inflation, which becomes
unbelievable steep within the last efold and results in an unbelievable steep decay of
correlator ${\mathscr C}_{\mathcal{RS}}$. For $n\gtrsim 56.6$, in particular, both
${\mathscr C}_{\mathcal{RS}}$ and ${\mathscr P}_{\mathcal S}$ become negligible\footnote{
If we define a parameter
\begin{equation}\label{eq: 51}
\alpha=\frac{{\mathscr P}_{\mathcal S}}{{\mathscr P}_{\mathcal R}
+ {\mathscr P}_{\mathcal S}}
\end{equation}
to measure the relative contribution of the isocurvature fluctuations, $\alpha |_{n=56.
83847}\approx 3.7\times 10^{-10}$. Hence the constraint on isocurvature perturbations
from observations\cite{enqvist} is trivially matched and there is actually no observable
isocurvature perturbation spectrum after inflation in the present two-field model.}, and
${\mathscr P}_{\mathcal R}$ stops varying. Therefore, in the present two-field model,
the power spectrum of curvature perturbation on superhorizon scales is changeable with
respect to time. This is in contrast with the single field inflation
case where the curvature power spectrum remains approximately a constant after Hubble
crossing. Figure \ref{curvaturespectrum} gives the details of our numerical result for power
spectrum of the curvature perturbation, where the three curves respectively stand for the
complete curvature spectrum ${\mathscr P}_{\mathcal R} = \frac{k^3}{2\pi^2} \left(|{\mathcal
R}_1|^2 + |{\mathcal R}_2|^2 \right)$ and its two terms $\frac{k^3}{2\pi^2} |{\mathcal
R}_1|^2$ and $\frac{k^3}{2\pi^2} |{\mathcal R}_2|^2$.
We see that for $2\lesssim n \lesssim 30$ on superhorizon scales the spectrum of curvature
perturbation originates almost exclusively from the fluctuations of the initial curvature
perturbation along the inflationary trajectory. While for $n \gtrsim 30$, the
curvature-isocurvature correlation becomes important, the spectrum at the final stage of
inflation contains considerable contributions from the interactions with the isocurvature
modes. As a result, the curvature perturbation spectrum after inflation is slightly higher
than the curvature perturbation spectrum at Hubble crossing.

\begin{figure}[ht]
\begin{center}
\includegraphics{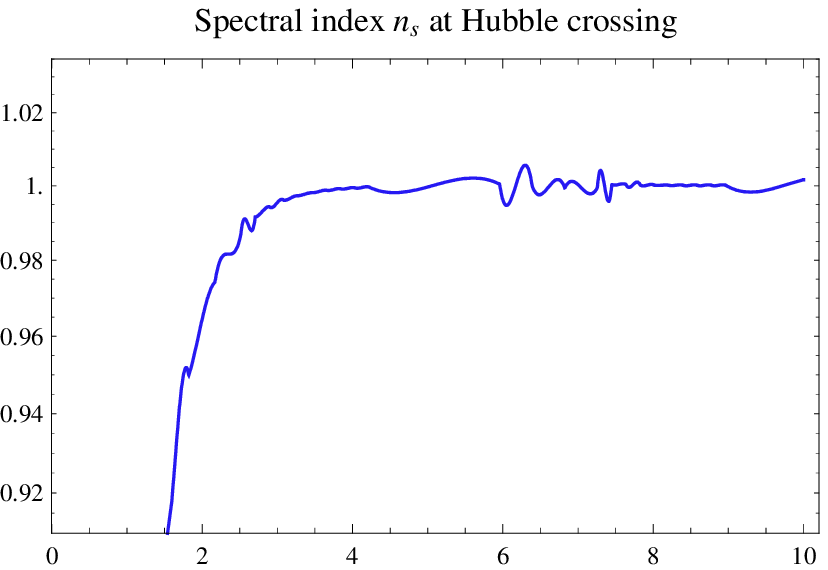}
\caption{\small Spectral index $n_s$ of the curvature power spectrum in the vicinity of
Hubble exit. The instant for COBE-like normalization corresponds roughly to $n = 2$,
where $n_s\approx 0.96$. However, the index running is sufficiently small so that for
$1.8\lesssim n \lesssim 50$, almost all values of $n_s$ are compatible with the observational
constraint.}
\label{index}
\end{center}
\end{figure}

The obtained power spectrum of the curvature perturbation in the present two-field model
is in good agreement with the COBE normalization of power spectrum\cite{mactavish}
\begin{equation}\label{eq: 52}
{\mathscr P}_{\mathcal R}^{\textrm{(COBE)}} \approx 3.96 \times 10^{-10}
\end{equation}
for $2\lesssim n \lesssim 30$. $n\approx 2$ is an instant approximately 55 efoldings
before the end of inflation, it is near the COBE normalization point. So $n\approx
2$ is a reasonable efold time corresponding to Hubble crossing. The scalar spectral
index of the curvature power spectrum
\begin{equation}\label{eq: 53}
n_s = 1 + {d \ln {\mathscr P}_{\mathcal R}}/{d \ln k}
\end{equation}
on superhorizon scales is also time-dependent, which has been calculated numerically with
partial results given in Figure \ref{index}. At the claimed Hubble crossing, $n_s|_{n \approx
2} \approx 0.96$, which is compatible with the best observational constraint available at
present\cite{seljak}, $n_s = 0.98 \pm 0.02$.

It follows from the above numerical investigation that, in the two-field model under
consideration, strong interactions exist between the curvature and isocurvature perturbations
during the final stage of inflation. Because the relative correlation coefficient
increases steadily from the assigned tiny value deep inside the Hubble radius to values
very close to \emph{one} at the end of inflation, a great deal curvature perturbation after
inflation originates from interactions between the curvature and isocurvature
perturbations on super-Hubble scales, not only from the quantum fluctuations of the initial
curvature perturbation along the classical trajectory. At this point, our model is somewhat
similar to the Double inflation with non-canonical kinetic terms and Roulette inflation
studied in \cite{langlois}. What distinguishes our model from the Double inflation with
non-canonical kinetic terms and Roulette inflation is that in the former the power spectrum
of isocurvature perturbation remains almost a constant during the most of inflationary epoch
while in the latter it decays rapidly after the Hubble radius crossing. This implies that
in our model the isocurvature perturbation continues to interact with the curvature
perturbation on superhorizon scales so that there is no an effective single field treatment
available for it\cite{bond}.

\section{Discussion}\label{sec:discuss}

We have established a multifield inflationary model on the large volume flux compactification
scheme in type IIB superstring orientifolds of $h^{(1,2)} >h^{(1,1)}$. The K\"ahler moduli
are categorized into volume modulus, heavy moduli and light moduli. Almost all of closed
string moduli emerging from the flux compactification are frozen at the potential minimum
in the spirit of KKLT mechanism in the large volume limit. However, this is not the case
for light K\"ahler moduli. The light K\"ahler moduli have very small masses, they are not
fixed by either the 3-form fluxes or nonperturbative effects in the effective superpotential,
and could be the driving force for the potential inflationary evolution of the universe in
its early history. To have a multifield inflation model, the superpotential is expected to
contain several exponent terms for the light K\"ahler moduli. Hence we have considered
the orientifolds of $h^{(1,1)}\geq 4$ and supposed that in our constructions the number of
light K\"ahler moduli is equal to or greater than two. In some sense, in this paper we have
proposed a plausible generalization of the better racetrack model\cite{quevedo061} in the
large volume approach\cite{quevedo051, quevedo052}.

The quantitative analysis of inflationary property in the resulting multifield inflationary
model has been done in a simplified two-field model where there are only two light K\"ahler
moduli with the axionic components being stabilized at the potential minimum also. This
assisted-like model is an extension of the K\"ahler moduli inflation model\cite{quevedo062}
to two-field inflation, which distinguishes itself by its kinetic Lagrangian that is neither
of canonical nor non-standard type studied in \cite{langlois}. Our investigation consists
of two steps. First, by some inevitable fine tuning in the initial conditions of the scalar
fields, we integrate the background equations numerically and obtain the inflationary
trajectory of the scalar fields. As a remarkable characteristic of the multifield model,
the trajectory in field space in our model are strongly bent roughly 50 efolds after
the Hubble crossing. The first slow-roll parameter is numerically calculated which turns
out to almost vanish for the first 60 efolds, implying that the scalar potential is very
flat along the trajectory. Second, we derive the evolution equations of the linear scalar
perturbations. These equations are numerically integrated on which we have calculated the
curvature, isocurvature spectra, and the correlation between them. Our numerical estimation
for the power spectrum of the curvature perturbation and the corresponding spectral index
is in good agreement with the COBE normalization and the WMAP observation dataset
\cite{mactavish, spergel}. In our model, the correlation between the curvature and
isocurvature perturbations is gradually strong on super-Hubble scales, so that the
curvature spectrum approximately thirty efolds after the Hubble exit is remarkably different
from the curvature perturbation spectrum at the Hubble exit, and there might be significant
non-Gaussianity in bispectrum and trispectrum of the primordial perturbations. The
investigation of the deviations from Gaussianity in power spectra in our two-field inflation
model is in progress.

\section*{Acknowledgments}
We would like to thank M.Li, J.X.Lu and Y.Wang for stimulating discussions. This work was
supported in part by CNSF-10375052 and the Startup Foundation of University of Science
and Technology of China and the Project of Knowledge Innovation Program (PKIP) of Chinese
Academy of Sciences.

%\newpage
%\section*{References}
%\bibliography{Reference}
%\bibliographystyle{utcaps}

\providecommand{\href}[2]{#2}\begingroup\raggedright\endgroup

\end{document}